\documentclass[a4paper,11pt]{article}

\usepackage[english]{babel}
\usepackage{setspace}

\usepackage[a4paper,tmargin=3.5truecm,bmargin=4truecm,rmargin=2.5truecm,
lmargin=3.2truecm,twoside,verbose=true]{geometry}
\usepackage{cancel,graphicx}
\usepackage{hyperref}
\usepackage{amsmath,amssymb,slashed}

\usepackage[amsmath, hyperref, thmmarks]{ntheorem}

\usepackage{xcolor} 
\usepackage[normalem]{ulem} 
\numberwithin{equation}{section}

\allowdisplaybreaks[1]
\newcommand\fabeta{{\tilde{f}}}
\newcommand\gabeta{{\tilde{g}}}

\newcommand\bbone{{\mathbb{I}}}

\newcommand\bbbone{\mathbb{I}}

\newcommand{\C}{\mathcal{C}}

\newcommand{\D}{\mathcal{D}}

\newcommand\DOM{\textup{Dom}}

\newcommand\dd{{\text{\textup{d}}}}

\renewenvironment{thebibliography}[1]
         {\section*{References}\frenchspacing\small
          \begin{list}{[\arabic{enumi}]}
         {\usecounter{enumi}\parsep=2pt\topsep 0pt
         \settowidth{\labelwidth}{[#1]}
         \leftmargin=\labelwidth\advance\leftmargin\labelsep
         \rightmargin=0pt\itemsep=1pt\sloppy}}{\end{list}}

\theoremsymbol{}
\theorembodyfont{\slshape}
\theoremheaderfont{\normalfont\bfseries}
\theoremseparator{}

\theorembodyfont{\upshape}
\theoremsymbol{\ensuremath{\blacklozenge}}

\theoremstyle{nonumberplain}
\theoremheaderfont{\scshape}
\theorembodyfont{\normalfont}
\theoremsymbol{\ensuremath{\blacksquare}}

\qedsymbol{\ensuremath{_\blacksquare}}
\theoremclass{LaTeX}

\begin{document}

\title{Quantum causality in $\kappa$-Minkowski and related constraints}
\author{Nicolas Franco$^a$, Kilian Hersent$^b$, Valentine Maris$^c$, Jean-Christophe Wallet$^b${\footnote{Author to whom any correspondence should be addressed.}}}

\date{}
\maketitle
\vspace*{-1cm}

\begin{center}
\textit{$^a$Namur Institute for Complex Systems (naXys) and Department of Mathematics, University of Namur, Namur, Belgium.}\\
\textit{$^b$IJCLab, Universit\'e Paris-Saclay, CNRS/IN2P3, 91405 Orsay, France.}  \\
\textit{$^c$ Fédération de Mathématiques FR3487, CentraleSupélec, 3 rue Joliot Curie, 91190 Gif-sur-Yvette, France .} \\
\textit{}\\
\bigskip
 e-mail:
\texttt{
    \href{mailto:nicolas.franco@unamur.be}{nicolas.franco@unamur.be}, %
    \href{mailto:kilian.hersent@universite-paris-saclay.fr}{kilian.hersent@universite-paris-saclay.fr}, %
    \href{mailto:valentine.maris@student-cs.fr}{valentine.maris@student-cs.fr}, %
    \href{mailto:jean-christophe.wallet@universite-paris-saclay.fr}{jean-christophe.wallet@universite-paris-saclay.fr}
    }\\[1ex]

\end{center}

\begin{abstract}
We study quantum causal structures in $1+1$ $\kappa$-Minkowski space-time described by a Lorentzian Spectral Triple whose Dirac operator is built from a natural set of twisted derivations of the $\kappa$-Poincar\'e algebra. We show that the Lorentzian Spectral Triple must be twisted to accommodate the twisted nature of the derivations. We exhibit various interesting classes of causal functions, including an analog of the light-cone coordinates. We show in particular that the existence of a causal propagation between two pure states, the quantum analogs of points, can exist provided quantum constraints, linking the momentum and the space coordinates, are satisfied. One of these constraints is a quantum analog of the speed of light limit.
\end{abstract}


\vfill\eject
\section{Introduction.}\label{section1}

Causality is a mandatory property of any realistic physical theory. While some approaches to Quantum Gravity \cite{phenoQG} pertain to the commutative world and thus exploit the usual notion of causality, e.g.\ causal-set \cite{bomb} or causal fermions systems \cite{causf}, other approaches come along with a noncommutative geometry framework \cite{connebook, connemarcol} which necessarily emerges at some step, mostly in term of noncommutative (quantum) space. These should of course incorporate one suitable notion of causality compatible with the noncommutative structure which can be called quantum causality. But it appears that different notions of causality can be defined in a noncommutative framework. For reviews, see e.g.\ \cite{cestbesnard, cestbesnardbis}.\\

Lorentzian noncommutative geometry \cite{eck-franc-copern} offers an appealing way to equip a quantum space-time with causal structures through a well defined notion of causality \cite{franc-eps2012}. This latter coincides with the usual causality at the commutative (low energy) limit. Recall that Lorentzian noncommutative geometry has been developed to accommodate Lorentzian signature in the standard framework of noncommutative geometry \cite{connebook, connemarcol}, the latter being rooted to a Riemanian framework. The central object is the so called Lorentzian Spectral Triple which, roughly speaking, is an adaptation of the celebrated Spectral Triple \cite{connebook} of the standard noncommutative geometry, whose related Dirac operator permits one to define a natural analog of the geodesic distance through the Connes spectral distance. Its construction on different noncommutative spaces has been considered in \cite{Ioch}-\cite{jcwrmp}. \\

As far as the Lorentzian Spectral Triple is concerned, the Dirac operator acts as a metric and is actually rigidly linked to its own notion of quantum causality. Quantum causality equipping almost commutative manifolds  has been considered in \cite{franc2014, francepst2015}. A related Zitterbewegung of a Dirac fermion has been evidenced in \cite{eckst2017}. The case of ``quantum Minkowski space-time'', which is nothing but the Moyal space equipped with a Minkowski metric, has been examined in \cite{francwal2016}. In particular, it has been shown that causal structures can exist in this quantum space suggesting that causal structures need not breakdown at the Planck scale, contrary to a common belief \cite{ameli-ellis, brukn}. \\

Quantum causality on $\kappa$-Minkowski space-time \cite{jurek}-\cite{jurek-rev}, one of the most promising quantum spaces in view of its possible role in Quantum Gravity{\footnote{The $\kappa$-Minkowski space-time can be defined as the dual of a subalgebra of the $\kappa$-Poincar\'e algebra involving the deformed translations, and may be viewed as the universal enveloping algebra of the Lie algebra of coordinates $[x_0,x_i]=\frac{i}{\kappa}x_i$, $[x_i,x_j]=0$, where $\kappa>0$. Recall that its "quantum symmetries" are coded by the $\kappa$-Poincar\'e algebra
which realises Doubly Special Relativity \cite{Lukierski_2003, Amelino_Camelia_2002a} and Relative Locality \cite{reloc}. See e.g.\ \cite{jurek-rev} for a general review.  }}, was first considered in \cite{Neves_2010}. Using some assumptions, it was found that the light-cone of the $\kappa$-Minkowski spacetime has a "blurry" region of 
Planck-length thickness which does not depend on the distance between any two events, which thus definitely would not produce detectable effects.\\
However, the authors of \cite{Mercati_2018a, Mercati_2018b} noticed that a part of the analysis in \cite{Neves_2010} depends on the basis used to describe the $\kappa$-Poincar\'e algebra. To bypass the base dependance problem, they defined a reasonable notion of Pauli-Jordan function from commutation relations between scalar fields. The resulting light-cone is still blurry but its width now depends on the square root of the distance between any two events. In particular, for cosmological distances, the effect due to blurriness is increased by $30$ orders of magnitude compared to the one found in \cite{Neves_2010}, which however is still not sufficient to give rise to detectable effects in the present experiments.\\

The above two analysis mainly exploit properties connected to field theories to model possible causal structures which may exist the $\kappa$-Minkowski space-time. Alternatively, a systematic description of a causal structure for the $\kappa$-Minkowski space-time, independently of any field theoretic consideration, can be directly extracted from any Lorentzian spectral triples modeling this quantum space. A determination of (hopefully typical) effects stemming from this type of causal structure, as done for the above two works is clearly worth to be carried out for physical purpose. However, the construction of a related Lorentzian Triple is not an easy task{\footnote{This is already the case in a purely Riemanian context.}}. Besides, different choices for the Dirac operator are possible, each leading {\it{a priori}} to its own causal structure, so that one legitimate question is to determine if these causal structures may share common properties, irrespective to the choice of a Dirac operator.\\

The corresponding exploration has been initiated in \cite{francwal22} where a particular Lorentzian Spectral Triple for $\kappa$-Minkowski in $1+1$ dimensions has been considered, the low dimensional situation being sufficient at this stage of investigation to tackle relevant properties, thus leaving aside additional dimensional complications. There, it was shown in particular that $\kappa$-Minkowski space-time admits specific causal relations, which are however ruled by a general constraint which is a quantum analog of the traditional speed of light limit. \\

The purpose of this paper is to pursue the above exploration of quantum causality in $\kappa$-Minkowski space-time, still staying in $1+1$ dimension. The Dirac operator defining the present Lorentzian Triple evades a problem faced by the one used in \cite{francwal22} in that its commutative ($\kappa\to\infty$) limit now coincides with the usual Dirac operator of the commutative Minkowski space-time, which clearly improves the approach of \cite{francwal22}. The price to be payed is that the present Dirac operator is built from twisted derivations. Twisted derivations show up in various instance, as explained below, and reduce to the usual derivations, say $\partial_\mu$, at the commutative limit. As a consequence of the appearance of twisted derivations in the present analysis, we will make use of {\it{twisted}} versions of Lorentzian Spectral Triples. We find however useful to recall the main features underlying the untwisted Lorentzian Spectral Triples and then list the few modifications arising in the twisted case. The mathematical details relevant for the notion of causality described by untwisted triples can be found in \cite{franc-eps2012, franc2014}. A more complete mathematical framework in the twisted case will be published elsewhere.\\
Recall that a Lorentzian Spectral Triple is defined by the following set of data:
\begin{equation}
    \big\{\mathbb{A},\ \widetilde{\mathbb{A}},\ \pi,\ \mathcal{H},\ D,\ \mathcal{J}  \big\}
    \label{data1},
\end{equation}
in which $\mathbb{A}$ is an involutive non-unital (pre-C$^*$) algebra, $\widetilde{\mathbb{A}}$ is a preferred unitalization of $\mathbb{A}$ admitting this latter as an ideal, $\mathcal{H}$ is a Hilbert space, $\pi$ is a faithful $*$-representation of both $\mathbb{A}$ and $\widetilde{\mathbb{A}}$ on the algebra of bounded operators on $\mathcal{H}$ denoted by $\mathcal{B}(\mathcal{H})$, $D$ is the (unbounded) Dirac operator with domain $\text{Dom}(D)$ dense in $\mathcal{H}$ and the operator $\mathcal{J}\in\mathcal{B}(\mathcal{H})$ is the so-called fundamental symmetry. The latter serves to relate Hilbert space with positive definite product $\langle\cdot\ ,\ \cdot\rangle$ to Krein space with indefinite inner product $(\cdot\ ,\ \cdot)_\mathcal{J}$ through 
\begin{equation}
    (\cdot\ ,\ \cdot)_\mathcal{J}=\langle\ \cdot\ , \mathcal{J}\ \cdot\rangle.
    \label{kreinproduitgeneric}
\end{equation}
 Notice that the introduction of a unital algebra $\widetilde{\mathbb{A}}$ is just needed for technical purpose. Recall that a unital algebra could not support a notion of causality. Indeed, a unital algebra models a compact space on which no causality can be defined.\\
The above data \eqref{data1} must obey the following conditions{\footnote{Note that the set of data \eqref{data1} is requested to satisfy a supplementary condition \cite{franc-eps2012, franc2014} which is given by $\pi(a) (1 + \langle D \rangle^2 )^{-\frac{1}{2}} \in \mathcal{K}(\mathcal{H})$ for any $  a \in {\mathbb{A}}$ where $\mathcal{K}(\mathcal{H})$ denotes the algebra of compact operators on $\mathcal{H}$ and $\langle D\rangle^2
    := \frac{1}{2} (D^\ddag D + D D^\ddag) $. This condition will not be needed in the ensuing analysis.}}:
\begin{align}
    & \forall a \in \widetilde{\mathbb{A}},\ [D, \pi(a)] \in \mathcal{B}(\mathcal{H}),
    \label{cond1}\\
    & \mathcal{J}^2 = 1,\ \ 
    \mathcal{J}^\ddag = \mathcal{J},\ \ 
    [\mathcal{J}, \pi(a)] = 0, \ \forall a\in\widetilde{\mathbb{A}}
    \label{cond2},\\
    & D^\ddag \mathcal{J} = - \mathcal{J} D,
    \label{cond3}
\end{align}
where $D^\ddag$ corresponds to the adjoint of $D$ for the Hilbert product of $\mathcal{H}$, the last condition \eqref{cond3}, which must hold on $\text{Dom}(D)=\text{Dom}(D^\ddag)$, insures that $D$ is self-adjoint w.r.t the indefinite inner product $(.\ ,\ .)_\mathcal{J}$ \eqref{kreinproduitgeneric}.

In order to insure a Lorentzian-type signature, equations \eqref{cond1}-\eqref{cond3} must be supplemented \cite{franc-eps2012, franc2014} by
\begin{equation}
    \mathcal{J} = -N[D, \mathcal{T}],\ \ 
    \label{cond6},
\end{equation}
in which $N$ is a positive element of $\widetilde{\mathbb{A}}$ to be determined and
$\mathcal{T}$ is a self-adjoint operator with dense domain in $ \mathcal{H}$, such that $\text{Dom}(D)\cap\text{Dom}(\mathcal{T})$ is dense in $\mathcal{H}$ in obvious notations and verifying 
\begin{equation}
    (1 + \mathcal{T})^{-\frac{1}{2}}
    \in \widetilde{\mathbb{A}}
    \label{cond5}.
\end{equation}
The positive operator $\mathcal{T}$ is nothing but the (noncommutative) analog of a global time function.\\

In the framework of Lorentzian Spectral Triples, an essential building block of the related (quantum) causality is provided \cite{franc-eps2012} by the causal cone $\mathcal{C}$. It is defined as the convex cone of hermitian elements $a\in\widetilde{\mathbb{A}}$ fulfilling 
\begin{equation}
  \langle\psi ,\mathcal{J}[\mathcal{D},\pi(a)]\psi\rangle\le0,  \label{condcone1}
    \end{equation}
for any $\psi\in\mathcal{H}$. Then, the causal relation between pure states, say $\omega$ and $\eta$, which can be viewed as the quantum (noncommutative) analogs of points (i.e.\ events in space-time), is defined as the following partial order relation
\begin{equation}
\forall \omega,\eta\in\mathfrak{S}(\widetilde{\mathbb{A}}),\ \omega\preceq \eta\iff \forall a\in\mathcal{C},\ \omega(a)\le\eta(a),\label{cestcausal2bis}
\end{equation}
where $\mathfrak{S}(\widetilde{\mathbb{A}})$ denotes the space of states of the unitalized algebra $\widetilde{\mathbb{A}}$. Note that \eqref{cestcausal2bis} may be extended to any states of $\widetilde{\mathbb{A}}$. Non-pure states may be interpreted as kind of wave packets so that causal relations between these objects can even be defined within this framework. As an important remark, it must be pointed out that the above quantum causality coincides with the usual causality at the commutative limit. It can be stressed that the above description stems from an algebraization of the usual causality prevailing on usual (globally hyperbolic) manifolds. This is summarized in Appendix \ref{ApendixA}.\\

One simple observation is that Lorentzian Spectral Triples can be viewed as (analogs of) pseudo-Euclidean versions of the Connes Spectral Triples which are rigidly rooted to an Euclidean/Riemanian framework \cite{connebook, connemarcol}. In these latter framework, as well as in the standard version of Lorentzian Spectral Triples summarized above, one common feature among others is that the commutators, such as those appearing in \eqref{cond1}, \eqref{cond2}, \eqref{cond6} and \eqref{condcone1}, retain their usual expression, namely one has formally $[a,b]=ab-ba$ in obvious notations.\\

But twisted Spectral Triples in Euclidian/Riemanian framework also appeared in various areas of the literature, e.g.\ within Type III operator algebras \cite{connemosco} or in relation with quantum groups, while they are essential in the construction of alternative versions of the Standard Model description "{\it{\`a la Connes}}" \cite{landi1}-\cite{martinet}. Twisted Spectral Triples for $\kappa$-Minkowski space in an Euclidean set-up have also been considered in \cite{Matassa2012, matasabis}. The salient feature of these triples is that the Dirac operator is built from twisted derivations, i.e.\ maps obeying a twisted Leibniz rule. This turns (most of) the commutators into twisted commutators, i.e.\ one now has formally $[a,b]=ab-\rho(b)a $ where $\rho$ is now some automorphism related to the twisted Leibniz rule.\\

In this paper, the Lorentzian Spectral Triples used to characterize properties of the causal structures prevailing on the 2-dimensional $\kappa$-Minkowski space-time will involve a Dirac operator built from the twisted derivations mentioned just above. This Dirac operator coincides with the usual Dirac operator at the $\kappa\to\infty$ (commutative) limit, thus evading the problem faced by the Dirac operator used in \cite{francwal22}. Recall that the above twisted derivations belong to the so-called deformed translations, a Hopf subalgebra of the $\kappa$-Poincar\'e algebra and have been used to generate the twisted differential calculus leading to the construction of the $\kappa$-Poincar\'e invariant gauge theories  on $\kappa$-Minkowski \cite{MW20, MW20-bis, zereviou}. According to the above discussion, the commutators appearing in \eqref{cond1}, \eqref{cond2}, \eqref{cond6} and \eqref{condcone1} will be twisted. Hence the Lorentzian triples may be viewed {\it{informally}} as pseudo-euclidean versions of the Spectral Triples of \cite{Matassa2012, matasabis}.\\

The paper is organised as follows. The Section \ref{section2} collects the properties characterizing the algebra, its unitalization, the fundamental symmetry together with the Dirac operator, which will be the elements common to the two Lorentzian triples considered in the present analysis. \\
Subsection \ref{section22} presents the essential properties of the representation of the algebra used in this analysis. \\
In Subsection \ref{section23}, we obtain the twisted conditions characterizing the Lorentzian Spectral Triple. It is shown that the twist appearing in these relations is directly related to the twist linked to the twisted derivations used to define the Dirac operator. \\
In Subsection \ref{section333}, we consider the conditions characterizing the causal cone and exhibit particular classes of functions, beyond the usual time function. Among these functions, a family bearing some similarity with the light-cone coordinates is exhibited.\\
In Section \ref{sectioncontraint}, we show that these functions are rigidly linked to quantum constraints on the momentum and space coordinate which represent necessary conditions for a causal propagation between pure states to exist. Causal functions of light-cone-type coordinates give rise in particular to a quantum analog of the speed of light limit. A sufficient condition for a continuous causal evolution between two pure states is examined. \\
In Section \ref{decadix}, we summarize and discuss our results.

\section{Algebra, fundamental symmetry and Dirac operator.}
\label{section2}
To describe conveniently the $\kappa$-Minkowski space-time, a natural framework for the relevant algebraic structures is provided \cite{PW2018, DS} by the convolution algebra for the two-dimensional affine group $\mathcal{G}=\mathbb{R}\ltimes\mathbb{R}$ \cite{khalil} and its related group algebras. This is the route we will follow in this paper. In this section, we will collect the relevant material. For the corresponding mathematical details, see \cite{PW2018, DS}. For the properties characterising the twisted derivations leading to the Dirac operator, see \cite{MW20, MW20-bis}.

\subsection{The associative algebras modeling the \texorpdfstring{$\kappa$}{kappa}-Minkowski space-time.}
\label{section21}
A  suitable choice \cite{francwal22} for the non-unital ${}^*$-algebra $\mathbb{A}$ involved in the Lorentzian Spectral Triples is
\begin{equation}
  \mathbb{A}_x
  =
  \Big( \mathfrak{E} \big( \mathbb{R}, \mathfrak{E}(\mathbb{R}) \big),\ 
  \star,\ \dag \Big)
  \label{algebrenx},
\end{equation}
where the associative product $\star$ and involution $\dag$ are defined by \cite{PW2018}, \cite{DS}
\begin{equation}
(f\star g)(x_0,x_1)=\int \frac{dp_0}{2\pi}dy_0\ e^{-iy_0p_0}f(x_0+y_0,{x_1})g(x_0,e^{-p_0/\kappa}{x_1})  \label{starpro-2d}.
\end{equation}
\begin{equation}
    f^\dag(x_0,x_1)= \int \frac{dp_0}{2\pi}dy_0\ e^{-iy_0p_0}{\bar{f}}(x_0+y_0,e^{-p_0/\kappa}{x_1})\label{invol-2d},
\end{equation}
and $\mathfrak{E}(\mathbb{R},\mathfrak{E}(\mathbb{R}))$ denotes the space of functions whose analytic continuation in the first variable $x_0$ is an entire function on $\mathbb{C}$ of exponential type{\footnote{A function $f:z\in\mathbb{C}\to f(z)$ is of exponential type if it satisfies an exponential bound, that is, if one has $|f(z)|\le K_1e^{K_2|\Im{z}|}$, where $K_1$ and $K_2$ are {\it{positive}} real constants.}} with values in the space of analytic functions of exponential type in the variable $x_1$. \\
Notice that the subscript $x$ in $\mathbb{A}_x$ \eqref{algebrenx} indicates the type of variables to be used. In the course of the analysis, it will be convenient to use Fourier transformed functions in the first variable, i.e.\ $(p_0,x_1)$, as well as Fourier transformed functions in both variables, i.e.\ $(p_0,p_1)$. The corresponding algebras will be denoted respectively by $\mathbb{A}_{px}$ and $\mathbb{A}_{p}$ as in ref.\ \cite{francwal22}. Since $\mathbb{A}_x:=\mathcal{F}^{-1}\mathbb{A}_p$, one infers that \cite{francwal22}
\begin{equation}
    \mathbb{A}_p = C^\infty_c(\mathbb{R}, C^\infty_c(\mathbb{R}))\label{algebrenp}
\end{equation}
with obvious corresponding changes in the expression for the associative product and involution to be recalled below, which is the algebra of smooth functions in the variables $p_0$ and $p_1$ with compact support for $p_0$ and taking their values in the space of smooth functions of $p_1$ with compact support. Recall that this algebra is dense in the group $C^*$-algebra \cite{dana} for the 2-d affine group characterizing the $\kappa$-Minkowski space-time \cite{PW2018, DS}. The natural product equipping \eqref{algebrenp} is nothing but the convolution product $\circ$ defining the convolution algebra of the affine group $\mathcal{G}=\mathbb{R}\ltimes\mathbb{R}$. For mathematical details on convolution algebras and algebras of locally compact groups, see e.g.\ \cite{dana}. Here, the convolution product is linked to the star-product \eqref{starpro-2d} through
\begin{equation}
\mathcal{F}(f\star g)=\mathcal{F}f\circ\mathcal{F}g\label{star-pro},
\end{equation}
stemming from the Weyl quantization map{\footnote{The Weyl quantization map is defined as $Q(f):=\pi(\mathcal{F}f)$ for any $f\in\mathbb{A}_p$ where $\pi$ is a non-degenerate $*$-representation of the group algebra for $\mathcal{G}$. To obtain \eqref{star-pro} , simply use the fact that $\pi$ is non-degenerate and a $*$-morphism of algebra. For a physics-oriented review on star-product construction, see \cite{zereviou} and references therein.}} \cite{PW2018}, \cite{DS}. One easily obtains
\begin{equation}
(\mathcal{F}f\circ \mathcal{F}g)(p_0,p_1)=\int\ dq_0dq_1e^{\frac{q_0}{\kappa}}\ \mathcal{F}f(q_0,q_1)\mathcal{F}g(p_0-q_0,e^{\frac{q_0}{\kappa}}(p_1-q_1))\label{convol-fin},
\end{equation}
\begin{equation}
\mathcal{F}f^\ddag(p_0,p_1)={\overline{\mathcal{F}f}}(-p_0,-p_1e^{\frac{p_0}{\kappa}})e^{-\frac{p_0}{\kappa}}\label{invol-fin},
\end{equation}
where \eqref{invol-fin} is the counterpart of \eqref{invol-2d}, ${\overline{\mathcal{F}f}} $ denotes the complex conjugation of $ {{\mathcal{F}f}}$ and the left Haar measure has been used in \eqref{convol-fin}. In the same way, the corresponding algebra for the variables $(p_0,x_1)$ obtained by partial Fourier transform is given by
\begin{equation}
\mathbb{A}_{px}=\mathcal{C}_c^\infty(\mathbb{R}, \mathfrak{E}(\mathbb{R})),\label{algebrenpx}
\end{equation}
with corresponding associative product
\begin{equation}
(\fabeta \star \gabeta)(p_0,x_1) = \int \dd q_0\; \fabeta(q_0,x_1)\, \gabeta(p_0-q_0, e^{-q_0/\kappa} x_1),
\end{equation}
where partially Fourier transformed functions are denoted by $\tilde{f}(p_0,x_1)$.\\

The Alexandroff unitalization of $\mathbb{A}$ is as in ref.\ \cite{francwal22}. Indeed, using the same notations as above, the unital algebras in the different set of variables, $\widetilde{\mathbb{A}}_x$, $\widetilde{\mathbb{A}}_p$ and $\widetilde{\mathbb{A}}_{px}$ are obtained by supplementing their respective non-unital counterpart by the following unit functions expressed in the suitable set of variables
\begin{equation}
\bbbone(x_0,x_1)=1,\ \bbbone(p_0,p_1)=(\delta(p_0),\delta(p_1)),\   \bbbone(p_0,x_1)=\delta(p_0)\ \ \label{unities}.
\end{equation}

\subsection{Fundamental symmetry and Dirac operator.}

In the present paper, the operator defining the fundamental symmetry is given by
\begin{equation}
    \mathcal{J}=i\gamma^0 \otimes \bbbone_3 \label{symfundam},
\end{equation}
where $\gamma^0$ denotes one of the 2-dimensional gamma matrices. Our conventions for the gamma matrices are: 
\begin{equation}
 \gamma^0 = \left(\begin{matrix}0 & i\\i&0\end{matrix}\right),\ \  \gamma^1 = \left(\begin{matrix}0 & -i\\i&0\end{matrix}\right)\nonumber. 
\end{equation}.\\

To define the Dirac operator equipping each of the Lorentzian Spectral Triples, we start from the following family of twisted derivations of the algebra{\footnote{Our conventions are those of \cite{MW20, MW20-bis}. }} \cite{MW20, MW20-bis}:
\begin{equation}
    X_0=\kappa(1-\mathcal{E}),\ \ X_1= P_1,
\label{2d-derivation}
\end{equation}
with $[X_0,X_1]=0$, where 
\begin{equation}
   \mathcal{E}=e^{-P_0/\kappa}.
\end{equation}
Recall that $(\mathcal{E},P_0,P_1)$ are the primitive elements of a Hopf subalgebra of the 2-dimensional $\kappa$-Poincar\'e algebra \cite{majid-ruegg}. This subalgebra is often called the deformed translation algebra.\\

The actions of $\mathcal{E}$ and $P_\mu$ ($\mu=0,1 $) on the algebra modeling the $\kappa$-Minkowski space-time are given by
\begin{equation}
    (\mathcal{E}\triangleright a)(x_0,x_1)=a(x_0+ \frac{i}{\kappa},x_1),\ 
    (P_\mu\triangleright a)(x_0,x_1)=-i\partial_\mu a(x_0,x_1)\label{cestlaction},
\end{equation}
for any function $a\in\mathbb{A}_x$. Note that higher dimensional versions of the above abelian Lie algebra of twisted derivations play a preeminent role in the construction of $\kappa$-Poincar\'e invariant gauge theory on $\kappa$-Minkowski space-time \cite{MW20}.\\
The twisted derivations \eqref{2d-derivation} satisfy the following twisted Leibniz rule
\begin{equation}
X_\mu(f\star g)=X_\mu(f)\star g+ (\mathcal{E}\triangleright f)\star X_\mu(g),\label{tausigleibniz}
\end{equation}
for any $f,g\in\mathbb{A}_x$. It must be stressed that they are {\it{not}} real derivations since one has
\begin{equation}
(X_\mu(f))^\dag=-\mathcal{E}^{-1}\triangleright (X_\mu(f^\dag))\ne X_\mu(f^\dag), \label{laformule}
\end{equation}
while one can easily check that they reduce to the usual (real) derivations $P_\mu$ at the commutative ($\kappa\to\infty$) limit. \\

Now, we define the Dirac operator as
\begin{equation}
  \D =-i\gamma^\mu X_\mu \otimes \bbbone_3 =\left(\begin{matrix}0 &X_-\\X_+ &0\end{matrix}\right) \otimes \bbbone_3:=D\otimes\bbbone_3\label{diracoper},
\end{equation}
where
\begin{equation}
    X_\pm=X_0\pm X_1\label{lightconeder},
\end{equation}
where $X_0$ and $X_1$ are given by \eqref{2d-derivation}. \\

Eqn.\ \eqref{diracoper} is not sufficient to fully define the Dirac operator. Indeed, one still has to determine on which Hilbert space \eqref{diracoper} acts. Two natural options corresponding to two different choices for the representation of the algebra of the Lorentzian Triple \eqref{data1} come into play. One is given by the unitary irreducible representations of $\mathcal{G}$, classified a long ago in \cite{cccp1}. The other one is the well known GNS representation. \\
In the rest of this paper, we will consider only the first option whose main properties will be recalled in the next section. In this respect, the domain of the operator $D$ appearing in \eqref{diracoper} will be chosen as $\DOM(D) = C^\infty_c(\mathbb{R}) \otimes \mathbb{C}^2$ which is dense in the Hilbert spaces \eqref{hilbert-final} defined below. The domain of the Dirac operator is then $\DOM(\mathcal{D}) = C^\infty_c(\mathbb{R}) \otimes \mathbb{C}^6$. \\

\noindent From now on, we set $\kappa=1$.

\section{Twisted conditions for the Lorentzian Triple.}\label{newsection3}
\subsection{Operators and unitary representations of the affine group.}\label{section22}

A convenient representation which can equip \eqref{data1} is obtained by starting from the unitary irreducible representations of the affine group $\mathcal{G}$ \cite{khalil, cccp1}, 
\begin{equation}
 \pi_U:\mathcal{G}\to \mathcal{B}(L^2(\mathbb{R},ds)),
 \end{equation}
 where the non-trivial one are given by
 \begin{equation}
 (\pi_{U\pm}(p_0,p_1)\phi)(s)=e^{\pm ip_1e^{-s}}\phi(s+p_0),\label{gelf1}
\end{equation}
for any $\phi\in L^2(\mathbb{R},ds)$ ($ds$ denotes the usual Lebesgue measure) while all the other unitary irreducible representations are 1-dimensional. \\
Accordingly, the non-degenerate {\it{bounded}} $^*$-representations of the group algebra linked to $\mathbb{A}_p$ \eqref{algebrenp} take the form
\begin{equation}
   \pi_\nu(f) :\mathbb{A}_p\to\mathcal{B}(L^2(\mathbb{R},ds)),\label{bounded-rep}
\end{equation}
\begin{equation}
(\pi_\nu(f)\phi)(s)=\int dp_0dp_1\ e^{p_0-s}\mathcal{F}f(p_0-s,p_1)e^{ i\nu p_1e^{-s}}\phi(p_0),\ \ \nu=-1,0,1\label{rep_p}
\end{equation}
for any $\phi\in L^2(\mathbb{R},ds)$. \\
In what follows, we will borrow most of the notations of ref. \cite{francwal22}. Notice that the representations \eqref{rep_p} are also irreducible since the representations of the affine group $\mathcal{G}$ are. This property will be useful to generate a family of pure states later on. Notice also that $\hat{\pi}=\pi_+\oplus\pi_-$ is faithful. \\

We will need the expression for the representations $\pi_\nu$ \eqref{rep_p} in the other sets of variables. After standard computations, the corresponding expressions in the spatial variables $(x_0,x_1)$ and in the mixed variables $(p_0,x_1)$ are respectively given by
\begin{equation}
    (\pi_{\nu}(f) \,\phi)(s)  
    = \tfrac{1}{2\pi} \int \dd u \dd v\; f(v, \nu e^{-s}) \,e^{-i v (u-s)}\, \phi(u),
    \ \ \nu=-1,0,1
\label{reprerpnu}
\end{equation}
and
\begin{equation}
    (\pi_{\nu}(f) \,\phi)(s) = \int \dd u\; \fabeta(u-s, \nu e^{-s})\, \phi(u),\ \ \nu=-1,0,1
    \label{reprerpmixed} .
\end{equation}
Owing to the fact that we will use a 2-dimensional Dirac operator, we set 
\begin{equation}
\mathcal{H}_{-,0,+}:=L^2(\mathbb{R},ds)\otimes\mathbb{C}^2,\label{blockhilbert}
\end{equation}
and define the Hilbert space $\mathcal{H}$ involved in \eqref{data1} as
\begin{equation}
\mathcal{H}=\mathcal{H}_{+}\oplus\mathcal{H}_{0}\oplus\mathcal{H}_{-}\label{hilbert-final},
\end{equation}
together with the faithful representation $\pi$ involved in \eqref{data1}
\begin{equation}
\pi=(\pi_+\oplus\pi_0\oplus\pi_-)\otimes\bbone_2.\label{cestpi}
\end{equation}
Note that the blockwise action of $\pi$ on $\mathcal{H}$ is obvious from \eqref{hilbert-final} and \eqref{cestpi}.\\

The Hilbert product equipping $\mathcal{H}$ is defined for any $\Phi,\Psi\in\mathcal{H}$ by
\begin{equation}
    \langle \Phi,\Psi \rangle=\sum_{\nu=+,-,0}\langle \varphi^{(\nu)},\psi^{(\nu)} \rangle_{\mathcal{H}_\nu}\label{hilbertproduct},
\end{equation}
 with $\Phi=\oplus_\nu\varphi^{(\nu)}$ according to the decomposition of $\mathcal{H}$ \eqref{hilbert-final} (and similarly for $\Psi$), where $\langle .,.\rangle_{\mathcal{H}_\nu}$, $\nu=+,-,0$, is the usual Hilbert product on $L^2(\mathbb{R},ds)\otimes\mathbb{C}^2$ given by 
\begin{equation}
\langle \varphi^{(\nu)},\psi^{(\nu)} \rangle_{\mathcal{H}_\nu}=\int ds\ {\varphi^{(\nu)}}^\dag(s)\psi^{(\nu)}(s)\label{prodhilbertutile}.
\end{equation}
The Krein product related to $\mathcal{J}$ is defined by
\begin{equation}
    (\Phi,\Psi)_{\mathcal{J}}:=\langle \Phi,\mathcal{J}\Psi \rangle=\sum_{\nu=+,-,0}\langle \varphi^{(\nu)},i\gamma^0\psi^{(\nu)} \rangle_{\mathcal{H}_\nu}\label{kreinproduit},
\end{equation}
where we used \eqref{symfundam}.\\

At this point, some comments on the above representation are in order.\\
By using \eqref{reprerpnu}, one infers that
\begin{equation}
\pi_\pm(x_0)=-i\frac{d}{ds}, \ \ \pi_\pm(x_1)=m(\pm e^{-s}),\label{chatdeschro}
 \end{equation}
where the symbol $m(.)$ denotes the left multiplication by $\pm e^{-s}$, which are clearly related to the well known Schr\"odinger (self-adjoint) operators of quantum mechanics $P, Q$ acting on $L^2(\mathbb{R},ds)$. Indeed, set 
\begin{equation}
P=x_0,\ \ x_1=e^{-Q}, \label{chatdeschro1}
\end{equation}
from which a simple calculation yields $[P,Q]=-i$.\\
Besides, since $x_1=e^{-Q}$ is a positive operator, it follows that $\pi_+(x_1)$ (resp. $\pi_-(x_1) $) models the positive (resp. negative) part of the line while the origin corresponds to $\pi_0(x_1)$.

\subsection{Twisted conditions for the Lorentzian Spectral Triple.}
\label{section23}
Let us now check that conditions \eqref{cond1}-\eqref{cond3} and \eqref{cond6}, \eqref{cond5} are verified, some of them however being twisted as discussed at the end of the introduction.\\

By straightforward algebraic manipulations, one can verify that $\mathcal{J}^2=1,\ \ \mathcal{J}^\dag=\mathcal{J}$ and $[\mathcal{J},\pi(a)]=0$ for any $a\in\widetilde{\mathbb{A}}$ so that condition \eqref{cond2} is fulfilled by the fundamental symmetry \eqref{symfundam} and the representation $\pi$ \eqref{cestpi}.\\

To verify the remaining conditions, one has first to implement the action of the derivations involved in the Dirac operator $\mathcal{D}$ on the Hilbert space $\mathcal{H}$ \eqref{hilbert-final}. Detailled computations of equations \eqref{commut1}, \eqref{commut2} and \eqref{commut3} are presented in Appendix \ref{apendix2}. As $\mathcal{D}$ acts blockwise on $\mathcal{H}$, it is convenient to consider the separate actions of the derivations on the Hilbert spaces $\mathcal{H}_\nu$ \eqref{blockhilbert}. Now by using \cite{CPT}
\begin{equation}
    (\partial_0\phi)(s) = s\phi(s)
    \label{act1},
\end{equation}
for any $ \phi\in C^\infty_c(\mathbb{R})\subset\mathcal{H}_\nu$, which realizes the {\it{self-adjoint}} action of the derivation $P_0$ onto the Hilbert spaces $\mathcal{H}_\nu$, $\nu=-1,0,1$, one verifies that
\begin{equation}
    ([\partial_0,\pi_\nu(a)]\phi)(s)=(i\pi_\nu(\partial_0a)\phi)(s),\ \ \nu=-1,0,1\label{commut1}
\end{equation}
for any $a\in\mathbb{A}_{x}$ and any $ \phi\in C^\infty_c(\mathbb{R})\subset\mathcal{H}_\nu$. The commutateur occurring in \eqref{commut1} is untwisted, reflecting the fact that $\partial_0\sim P_0$ is an untwisted derivation. Note that the computation is most easily carried out by using the mixed variables. \\
In the same way, by using
\begin{equation}
    (X_0\phi)(s)=(1-e^{s})\phi(s)
    \label{act2}
\end{equation}
valid for any $\phi\in C^\infty_c(\mathbb{R})\subset\mathcal{H}_\nu$, one finds from \eqref{commut1}
\begin{equation}
   ([X_0,\pi_\nu(a)]_\mathcal{E}\phi)(s)=(\pi_\nu(X_0(a))\phi)(s) ,\ \ \nu=-1,0,1\label{commut2}
\end{equation}
for any $a\in\mathbb{A}_{x}$ and any $ \phi\in C^\infty_c(\mathbb{R})\subset\mathcal{H}_\nu$, where $[.,.]_\mathcal{E}$ denotes the {\it{twisted commutator}} defined by
\begin{equation}
    [X_\mu, \pi_\nu(a)]_\mathcal{E}
    :=X_\mu \pi_\nu(a) - \pi_\nu(\mathcal{E}(a)) X_\mu
    \label{deftwist-commut}.
\end{equation}
Finally, upon using
\begin{equation}
    (X_1\phi)(s)=i\nu e^s\partial_s\phi(s)\label{act222-text},
\end{equation}
for any $ \phi\in C^\infty_c(\mathbb{R})\subset\mathcal{H}_\nu$, a similar computation yields
\begin{equation}
  ([X_1,\pi_\nu(a)]_\mathcal{E}\phi)(s)=(\pi_\nu(X_1(a))\phi)(s),\ \ \nu=-1,0,1\label{commut3}
\end{equation}
for any $a\in\mathbb{A}_{x}$ and any $ \phi\in C^\infty_c(\mathbb{R})\subset\mathcal{H}_\nu$. \\

As indicated in the introduction, it turns out that the Lorentzian Spectral Triple \eqref{data1} satisfies twisted versions of the conditions \eqref{cond1}, \eqref{cond3} and \eqref{cond6}, reflecting the fact that the derivations $X_0$ and $X_1$ are twisted. Indeed, first observe that one has the following useful relations
\begin{align}
    (\mathcal{E}^{-1}X_0\mathcal{E}\phi)(s)
    &= (1 - e^s) \phi(s), &
    (\mathcal{E}^{-1}X_1\mathcal{E}\phi)(s)
    &= i \nu\partial_s (e^s\phi)(s)
    \label{form111},
\end{align}
which hold for any $\phi\in C^\infty_c(\mathbb{R})$, as it can be easily verified by a straightforward computation. Notice that the LHS of both expressions in \eqref{form111} has to be understood as representing the successive actions of $\mathcal{E}$, $X_\mu$ and $\mathcal{E}^{-1}$ on any function $\phi$ in the domain $\text{Dom}(D)$ defined above. Then, by using \eqref{form111} together with \eqref{kreinproduit} and defining the following twist
\begin{equation}
    \tau:=\mathcal{E}\otimes\bbone_2\label{twistetau}
\end{equation}
one infers\\
\begin{equation}
    (\phi,\tau^{-1}D\tau\psi)_{i\gamma^0}=-(D\phi,\psi)_{i\gamma^0}\label{hermit-cond-twist},
\end{equation}
for any $\phi,\psi\in C^\infty_c(\mathbb{R}) $ where $(.,.)_{i\gamma^0}$ can be straightforwardly read off from \eqref{kreinproduit}. This implies
\begin{equation}
    D^\ddag\mathcal=-{(i\gamma^0)}\tau^{-1}D\tau{(i\gamma^0)}\label{dirac-dag}
\end{equation}
where the symbol $\dag$ in the LHS corresponds to the adjoint operation w.r.t.\ the Hilbert product on $\mathcal{H}_\nu$, which can be interpreted as a twisted hermiticity condition. \\

Finally, owing to the blockwise action of the Dirac operator $ \mathcal{D}$ \eqref{diracoper} on the Hilbert space, one easily obtains the natural extension of the condition \eqref{cond3} as
\begin{equation}
    \mathcal{D}^\ddag=-\mathcal{J}\rho^{-1}\mathcal{D}\rho\mathcal{J},\label{hermitordu}
\end{equation}
where
\begin{equation}
    \rho=\tau\otimes\bbone_3.\label{cesttau}
\end{equation}
is the twist acting on the Dirac operator $\mathcal{D}$.\\

At this point, one comment is in order.\\
The use of twisted derivations $X_0$ and $X_1$ in the construction of the Dirac operator reflects itself into the representation of the action of these derivations on the relevant domain $C^\infty_c(\mathbb{R}) $, resulting in the appearance of the extra twist $\rho$ \eqref{cesttau}. Notice that $\rho$ is related to the modular twist $\sim\mathcal{E}$ rooted in the present description of the algebra for $\kappa$-Minkowski \cite{MW20}. The action of the later also appeared as twisted commutators in \eqref{commut2} and \eqref{commut3}. \\

As we now show, this will be also the case for the condition \eqref{cond1} which will translate into a twisted commutator. Indeed, by using \eqref{commut2} and \eqref{commut3}, one obtains
\begin{equation}
    ([D,\pi_\nu(a)]_\mathcal{E}\phi)(s)=(\pi_\nu(Da)\phi)(s),\ \ \nu=-1,0,1\label{commut-dirac}
\end{equation}
for any $a\in\mathbb{A}_{x}$ and any $ \phi\in C^\infty_c(\mathbb{R})$. But $\pi_\nu $ is bounded as recalled at the beginning of this subsection, see \eqref{bounded-rep}. Owing to the blockwise action of $\mathcal{D}$ \eqref{diracoper} on $\mathcal{H}$ \eqref{hilbert-final}, it follows that
\begin{equation}
    [\mathcal{D},\pi(a)]_\mathcal{E}\in\mathcal{B}(\mathcal{H}),\label{twisbounded}
\end{equation}
which is a twisted extension of the condition \eqref{cond1}. It could be easily verified that the replacement of the twisted commutator $[.,.]_\mathcal{E}$ by its untwisted counterpart would lead to an unbounded operator. Note that the twisted commutator \eqref{twisbounded} will replace its untwisted counterpart in the definition of the causal cone introduced in Section \ref{newsection3}.\\

The appearance of a twist in the commutator of the condition \eqref{cond1} can be expected by noticing that the present Lorentzian Spectral Triple bears common ingredients with the modular (twisted) Spectral Triple related to the Euclidean version of the $\kappa$-Minkowski space considered in \cite{Matassa2012}. Both have {\it{formally}} similar Dirac operators (the symbols are the same) while the representations are different, the GNS representation, $\pi_{GNS}$, being used in \cite{Matassa2012}. In the modular triple, the introduction of a twisted commutator of the form (in obvious notations) $[D,\pi_{GNS}(a)]_\mathcal{E}$ guaranties that this latter is a bounded operator. Changing the representation does not modify this feature which may be viewed as stemming from the fact that the derivations \eqref{2d-derivation} used in both case are twisted.\\

Finally, by combining \eqref{commut2} and \eqref{commut3} with $D$ in \eqref{diracoper}, one readily obtains
\begin{equation}
    [D,\pi_\nu(x_0)\otimes\bbone_2]_\mathcal{E}=\pi_\nu(D(x_0))=\begin{pmatrix}
   0&1\\1&0 \end{pmatrix},
\end{equation}
for $\nu=-1,0,1$ from which one deduces that
\begin{equation}
    [\mathcal{D},\mathcal{T}]_\mathcal{E}=-\mathcal{J}
\end{equation}
with the time function given by
\begin{equation}
    \mathcal{T}=\oplus_\nu(\pi_\nu(x_0)\otimes\bbone_2),
\end{equation}
so that the condition \eqref{cond6} holds true with $N=\bbone$.\\

\subsection{Functions of the causal cone.}\label{section333}
In the present analysis, we will use the notion of quantum causality considered and developed in \cite{franc-eps2012}-\cite{francwal2016}. In this framework, the causality should be regarded as a partial order relation (denoted by the symbol $\preceq$ in the following discussion) between states of the unitalized algebra $\widetilde{\mathbb{A}}$. \\

Recall, as already mentioned in the introduction, see eqn. \eqref{condcone1}, that in the standard (untwisted) Lorentzian Spectral Triples \cite{franc-eps2012} the causal cone $\mathcal{C}$ is defined as the convex cone of hermitian elements $a\in\widetilde{\mathbb{A}}$ fulfilling the condition $\langle\psi ,\mathcal{J}[\mathcal{D},\pi(a)]\psi\rangle\le0$ for any $\psi\in\mathcal{H}$ in which the commutator is the usual one. We will indicate the relevant set of variables by the corresponding subscript when necessary, e.g.\ $\mathcal{C}_{px}$ denotes the causal cone with functions expressed with the mixed variables.\\

However, as explained in the introduction, the occurrence of a twisted commutator in \eqref{twisbounded} leads to the natural modification of the definition of the causal cone \eqref{condcone1} as being the convex cone of hermitian elements $a\in\widetilde{\mathbb{A}}$, denoted by $\mathcal{C}$, which fulfills 
\begin{equation}
    \forall\psi\in\mathcal{H},\ \ \langle\psi ,\mathcal{J}[\mathcal{D},\pi(a)]_\mathcal{E}\psi\rangle\le0.
    \label{lecone}
\end{equation}
In other words, this formula involves the twisted commutator, introduced in Subsection \ref{section23}. Here, $\mathcal{D}$ and $\pi$ are still given by \eqref{diracoper} and \eqref{cestpi}. \\
Then, the causal relation between pure states is defined by the following partial order relation
\begin{equation}
\forall \omega,\eta\in\mathfrak{S}(\widetilde{\mathbb{A}}),\ \omega\preceq \eta\iff \forall a\in\mathcal{C},\ \omega(a)\le\eta(a),
    \label{cestcausal2}
\end{equation}
where $\mathfrak{S}(\widetilde{\mathbb{A}})$ denotes the space of pure states of the unitalized algebra $\widetilde{\mathbb{A}}$, which may be extended to any states of $\widetilde{\mathbb{A}}$. A very brief summary of the above framework is presented in Appendix \ref{ApendixA}.\\

From \eqref{lecone}, one easily infers that an element $a\in\widetilde{\mathbb{A}}_{px}$ belongs to the causal cone $\mathcal{C}$ whenever
\begin{equation}
    \langle\psi ,\pi_\nu(iX_\pm(a))\psi\rangle\ge0,\ \ \nu=-1,0,1\label{leconecausal}
\end{equation}
for any $\psi\in C^\infty_c(\mathbb{R})$, i.e.\ the operator $\pi_\nu(iX_\pm(a))$ defines a positive operator. This condition thus characterizes the causal cone relevant to our analysis.\\

Eqn. \eqref{leconecausal} can be rewritten as
\begin{equation}
    \langle \psi,\pi_\nu(iX_\pm(a))\psi\rangle
    = \int dsdu\ \big(i(1-e^{s-u})\tilde{a}(u-s,\nu e^{-s})
    \pm\partial_\beta\tilde{a}(u-s,\nu e^{-s})\big)\psi(u)\overline{\psi(s)}\ge0,\label{levraycone}
\end{equation}
where the symbol $\partial_\beta$ denotes the derivative w.r.t. the second variable, which must hold for any $\psi \in C^\infty_c(\mathbb{R})$. Eqn. \eqref{levraycone} permits one to determine the functions in the causal cone $\mathcal{C}_{px}$.\\

While the full characterization of the causal cone $\mathcal{C}$ is beyond the scope of this paper, it is instructive to look for some non trivial functions belonging to $\mathcal{C}$, showing by the way that $\mathcal{C}$ is not the empty set. \\

First, one observes that constant are obviously in $\mathcal{C}$. Next, by combining $iX_\pm(x_0)=1$ with \eqref{leconecausal}, one straightforwardly obtains 
\begin{equation}
\langle\psi ,\pi_\nu(iX_\pm(x_0))\psi\rangle=\langle\psi ,\psi\rangle\ge0,    
\end{equation}
implying that the time function $T:=x_0$ belongs to $\mathcal{C}$. \\
Now, consider the family of functions 
\begin{equation}
    a_\pm(x_0, x_1):=x_0\pm \lambda x_1
    \label{light-cone-coor}
\end{equation}
with $\lambda$ a real constant, which are somewhat reminiscent of coordinates of light-cone type. An elementary calculation yields 
\begin{equation}
    iX_\pm(a_+(x_0,x_1)) 
    = iX_\mp(a_-(x_0,x_1))
    = 1 \pm \lambda 
\end{equation}
which combined with \eqref{leconecausal} show that $a_\pm(x_0,x_1)$ \eqref{light-cone-coor} belongs to the causal cone if and only if $\lambda \in  [-1, 1]$. \\
Finally, consider the family of functions
\begin{equation}
    a(x_0,x_1)=h(x_0)+g(x_1) \label{generalcoord}
\end{equation}
where $h,g\in\mathbb{A}$. Eqn. \eqref{generalcoord} plugged into \eqref{levraycone} yields the following condition
\begin{equation}
    \int dv\ (iX_0(h))(v)\vert \mathcal{F}\psi(v)\vert^2\pm\int ds\ g^\prime(\nu e^{-s})\vert\psi(s) \vert^2\ge0\label{marechalcond},
\end{equation}
where $e^{i\partial_v}\int dv\ e^{-ipv}=\int dv\ e^pe^{-ipv}$ and $\int dv\ (e^{-i\partial_v}e^{-ipv})f(v)=\int dv\ e^{-ipv}e^{i\partial_v}f(v)$ for any $f\in\mathbb{A}$ have been used in the computation. Thanks to the Plancherel theorem, the condition \eqref{marechalcond} is obviously satisfied provided the functions $h$ and $g$ verify
\begin{equation}
   iX_0(h)=1,\label{marechal1}
\end{equation}
and
\begin{equation}
    \vert g^\prime(\nu e^{-s})\vert\le1.\label{marechal2}
\end{equation}
From \eqref{marechal1}, one infers that any function $a$, decomposing through \eqref{generalcoord}, must satisfy
\begin{equation}
    h(x_0+i)-h(x_0)=i,\label{marechal3}
\end{equation}
which, in view of the properties of the algebra \eqref{algebrenx} defining the Lorentzian Triple, indicates that $h\in\mathbb{A}$ must be a quasiperiodic entire function of exponential type. \\

Observe that the time function $T=x_0$ mentioned above fulfils this property. This solution is not unique. Pick for instance $h(x_0) = x_0 + C e^{2 \pi k x_0}$, where $k \in \mathbb{Z}$ and $C$ is some constant. It can be easily verified to be a solution of \eqref{marechal1}. Besides, eqn.\ \eqref{marechal2} is verified by any functions $g\in\mathbb{A}$ which does not vary faster than $g(x_1)=x_1$. \\
Note that in \cite{francwal22} based on a different Dirac operator, the corresponding function $g$ was found not to vary faster than a logarithm while the counterpart of $h$ in \eqref{generalcoord} was equal to
$h(x_0)=x_0$ (up to an unessential additive constant). This simply reflects the difference between the two Dirac operators.\\

\section{Quantum constraints from quantum causality.}
\label{sectioncontraint}
\subsection{Pure states.}
From now on, we will consider the set of pure states underlying the analysis in \cite{francwal22}. Recall that a pure state is the natural analog of a point in the noncommutative setting. The present set of pure states is defined by the family of vector states
\begin{equation}
    \omega_\pm^\Phi:\mathbb{A}\to\mathbb{C},\ \ \omega_\pm^\Phi(a)=\langle \Phi,\pi_\pm(a)\Phi\rangle\label{etatpur}
\end{equation}
for any $\Phi\in\mathcal{H}_\pm$ \eqref{blockhilbert} with $||\Phi||=1$. Recall that the pure state nature of \eqref{etatpur} stems from the irreducible property of the representation $(\pi_\pm,\mathcal{H}_\pm)$. \\

Eqn.\ \eqref{etatpur} expressed in the mixed variables $(p,x)$ takes the form
\begin{equation}
  \omega_\pm^\Phi({a})=\int\dd u \dd s\ 
  \tilde{a}(u-s,\pm e^{-s})\overline\Phi(s)\Phi(u),\label{statetwo}
\end{equation}
for any $\Phi\in\mathcal{H}_\pm$. Then, from \eqref{cestcausal2}, one concludes that the causality relation between two pure states in the family defined by \eqref{etatpur} is simply
\begin{equation}
   \omega_\pm^{\Phi_1} \preceq \omega_\pm^{\Phi_2} \iff \forall a\in\C_x,\ \ 
  \omega_\pm^{\Phi_1}(a) \leq \omega_\pm^{\Phi_2}(a)  ,
\end{equation}
where the RHS can be conveniently expressed with the help of the mixed variables $(p,x)$ as
\begin{equation}
 \int \dd s \dd u \ \tilde{a}(u-s,  \pm e^{-s}) \, \left[ \overline\Phi_2(s)\Phi_2(u) - \overline\Phi_1(s)\Phi_1(u) \right] \geq 0,
 \label{evol-macro}
\end{equation}
for any $\tilde{a}\in\C_{px}$. Alternatively, \eqref{evol-macro} expressed with the space-time variables $(x_0,x)$ takes the form
\begin{equation}
\int \dd s \dd u \dd v\ e^{-iv(u-s)}a(v,\pm e^{-s})\left[ \overline\Phi_2(s)\Phi_2(u) - \overline\Phi_1(s)\Phi_1(u) \right] \geq 0,\label{evol-macro-bis}
\end{equation}
for any ${a}\in\mathcal{C}_{x}$.\\

\subsection{Quantum constraints.}\label{subsectioncontraint}

It appears that the partial order relation \eqref{cestcausal2} leads to an interesting physical interpretation. Indeed, assume for instance 
\begin{equation}
   a(v,\pm e^{-s})=v \pm e^{-s},\label{lesvraislightconecoord}
\end{equation}
which is one light-cone-type coordinate \eqref{light-cone-coor} for $\lambda = 1$ and thus belongs to the causal cone. Then, the resulting first term in the LHS of \eqref{evol-macro-bis} can be written as
\begin{equation}
\int \dd s \dd u \dd v\ e^{-iv(u-s)}v\left[ \overline\Phi_2(s)\Phi_2(u) - \overline\Phi_1(s)\Phi_1(u) \right]= \int \dd v\ v(\vert(\mathcal{F}\Phi_2)(v) \vert^2-\vert(\mathcal{F}\Phi_1)(v)\vert^2 ),
\end{equation}
where
\begin{equation}
(\mathcal{F}\Phi)(v):=\int \dd u\ e^{-iuv}\Phi(u).
\end{equation}
By using standard manipulations on the Fourier transforms, one can write
\begin{equation}
 \int \dd v\ v\vert(\mathcal{F}\Phi)(v) \vert^2=\int \dd v\ \overline{(\mathcal{F}\Phi)(v)}(\mathcal{F}(-i\frac{d}{dv}\Phi))(v)=\int \dd v\ \overline{\Phi(v)}(-i\frac{d}{dv}\Phi)(v),
\end{equation}
where the rightmost equality stems from the Plancherel theorem, which simply expresses the expectation value of the Schr\"odinger operator $P = -i\frac{d}{ds}$ \eqref{chatdeschro1} in the state $\Phi$. Therefore, one can write
\begin{equation}
 \int \dd v\ v\vert(\mathcal{F}\Phi)(v) \vert^2=\langle \Phi\vert P\vert\Phi\rangle\label{expectP}.
\end{equation}
The second term in \eqref{evol-macro-bis} can be easily cast into the form
\begin{equation}
\int \dd s \dd u \dd v\ e^{-iv(u-s)}e^{-s}\left[ \overline\Phi_2(s)\Phi_2(u) - \overline\Phi_1(s)\Phi_1(u) \right]=\int \dd s\ e^{-s}\left[ \vert\Phi_2(s)\vert^2 - \vert\Phi_1(s)\vert^2 \right]\label{mia11},
\end{equation}
which, owing to \eqref{chatdeschro1}, can be interpreted as the difference between the expectations of the operator $X=e^{-Q}$ in the states $\Phi_1$ and $\Phi_2$, namely
\begin{equation}
\int \dd s \dd u \dd v\ e^{-iv(u-s)}e^{-s}\left[ \overline\Phi_2(s)\Phi_2(u) - \overline\Phi_1(s)\Phi_1(u) \right]=\langle\Phi_2\vert X \vert \Phi_2\rangle-\langle\Phi_1\vert X \vert \Phi_1\rangle\label{mia1}.
\end{equation}
Putting all together, one concludes that a causal evolution from $\Phi_1$ to $\Phi_2$
can occur provided
\begin{equation}
    \langle \Phi_2 \vert P \vert \Phi_2\rangle
    - \langle \Phi_1\vert P \vert \Phi_1\rangle
    \ge \big\vert \langle \Phi_2 \vert X \vert \Phi_2 \rangle
    - \langle \Phi_1 \vert X \vert \Phi_1 \rangle \big\vert
    \label{CN-causal}
\end{equation}
holds true.\\
The condition \eqref{CN-causal} is one necessary condition for a quantum causal evolution between pure states (i.e. the quantum analogous of points/events) to exist. \\

At this point, some comments are in order.\\
First, one observes that it bears some similarity with the condition found in \cite{francwal22} based from the use of a different Lorentzian Spectral Triple to model the $\kappa$-Minkowski space-time. Indeed, denoting the variation of the expectation for $P$ in two states by $\delta\langle P\rangle$ and the corresponding variation for the relevant spatial coordinate by $\delta\langle X\rangle$, one obtains from \eqref{CN-causal}
\begin{equation}
    \delta\langle P\rangle\ge \vert\delta\langle X\rangle\vert\label{limitvitesse}
\end{equation}
which has the same structure than eqn.\ (126) of \cite{francwal22}
which was already interpreted as a quantum analogous of the classical speed of light limit. Simply recall that in a 2-dimensional Minkowski spacetime, two events are causally related if and only if variations of the spatial coordinate $\delta x$ and time coordinate $\delta t$ obey the inequality $\delta t\ge\vert \delta x\vert$. This property combined with \eqref{chatdeschro1} supports the above interpretation, with classical quantities replaced by expectation values. \\
Notice however that the present definition of the spatial coordinates differs from the one of ref.\ \cite{francwal22}, presently related to the canonical variable $Q$ \eqref{chatdeschro1} as $X=e^{-Q}$. This reflects the difference between the Dirac operators used in \cite{francwal22} and the present work.\\

Next, it can be observed that the constraint \eqref{limitvitesse} ensures that any function of the causal cone of the form given by \eqref{light-cone-coor} fulfils the condition \eqref{evol-macro-bis} since $|\lambda|\le1$. Other choices for $a$ of the form \eqref{generalcoord} would give rise to constraints of the generic form 
\begin{equation}
  \delta\langle h(P)\rangle\ge|\delta\langle g(X)\rangle | \label{mystery} 
\end{equation}
where $h$ and $g$ verify respectively \eqref{marechal1} and \eqref{marechal2}. The physical interpretation of these constraints is however unclear so far. It would be interesting to determine if they have some classical (low energy) imprint or if they are of purely quantum nature.\\

Finally, from the above analysis, one observes that any function in the causal cone, in addition to those having the splitted form \eqref{generalcoord}, generates a constraint stemming from the condition \eqref{evol-macro}, which should therefore result in a huge set of constraints. One other interesting issue would be to determine if there is a kind of hierarchy among these constraints. This requires a complete characterisation of the causal cone. Note that the above analysis and the above conclusions apply to a particular set of pure states. The complete set of pure states of the algebra considered in this analysis is unknown.\\

\subsection{Continuous causal evolution.}


It appears that one can establish a {{sufficient condition}} ruling the existence of a continuous causal evolution between these pure states. First,  by a mere use of the second fundamental theorem of analysis, eqn. \eqref{evol-macro} can be transformed into 
\begin{equation}
 \int \dd s \dd u \ \tilde{a}(u-s,  \pm e^{-s})\, \frac{d}{dt}  \left(  \overline\Phi(t;s)\Phi(t;u) \right) \geq 0,\label{evolstatescont}
\end{equation}
which must hold for any $\tilde{a}\in\C_{px}$, where now
$\Phi(t;s)$ models a stepwise evolution in $\mathcal{H}_\pm$ from $\Phi_1:=\Phi(t=1;s)$ to $\Phi_2:=\Phi(t=2;s)$ related to a continuous evolution from the state $\omega_\pm^{\Phi_1} $ to the state $\omega_\pm^{\Phi_2}$, labelled by a continuous parameter $t\in[1,2]$.\\

Now, consider \eqref{levraycone} and define
\begin{equation}
    F(u,s):= \nu e^{s} \widetilde{a}(u-s,\nu e^{-s}) \psi(u) \overline{\psi(s)}
    \label{grandF}.
\end{equation}
Since $\psi(s)\in C^\infty_c(\mathbb{R})$, one has by the Stokes theorem
\begin{equation}
   \int \dd s \dd u\ \left(\frac{\partial F(u,s)}{\partial s}+\frac{\partial F(u,s)}{\partial u}\right) =0.\label{stokes}
\end{equation}
Next, one simply computes
\begin{align}
    \frac{\partial F(u,s)}{\partial s}
    = &\ \nu e^{s} \widetilde{a}(u-s,\nu e^{-s}) \psi(u)\overline{\psi(s)}
    - \nu e^{s} \partial_a \widetilde{a}(u-s,\nu e^{-s}) \psi(u)\overline{\psi(s)} \nonumber\\
    &- \partial_\beta \widetilde{a}(u-s,\nu e^{-s}) \psi(u)\overline{\psi(s)}
    + \nu e^{s} \widetilde{a}(u-s,\nu e^{-s}) \psi(u)\overline{\psi^\prime(s)}
    \label{eqn1}, \\
    \frac{\partial F(u,s)}{\partial u}
    = &\ \nu e^{s} \partial_a \widetilde{a}(u-s,\nu e^{-s}) \psi(u)\overline{\psi(s)}
    + \nu e^{s} \widetilde{a}(u-s,\nu e^{-s}) \psi^\prime(u)\overline{\psi(s)},
    \label{eqn2}
\end{align}
in which $\partial_a$ and $\partial_\beta$ denote respectively the partial derivative w.r.t.\ the first and second variable for functions expressed with the mixed variables while $\psi^\prime(s) := \frac{d\psi(s)}{ds}$. By combining \eqref{stokes}, \eqref{eqn1} and \eqref{eqn2}, the second term in \eqref{levraycone} can be re-expressed as
\begin{equation}
\begin{aligned}
    \int \dd s \dd u\ &\partial_\beta\widetilde{a}(u-s,\nu e^{-s})\overline{\psi(s)}\psi(u) \\
    &= \int \dd s \dd u\ \nu e^{s}\widetilde{a}(u-s,\nu e^{-s}) \big( \psi(u)\overline{\psi^\prime(s)} + \psi^\prime(u) \overline{\psi(s)} + \psi(u)\overline{\psi(s)} \big)
    \label{eqn3}.
\end{aligned}
\end{equation}
Finally, inserting \eqref{eqn3} into \eqref{levraycone} and confronting the result with \eqref{evolstatescont}, one easily obtains a {\it{sufficient condition}} ruling the existence of a continuous causal evolution between states. \\

The sufficient condition can be stated as follows:\\
A continuous causal evolution from a pure state $\omega_\pm^{\Phi_1}$ to a pure state $\omega_\pm^{\Phi_2}$, i.e.\ $\omega_\pm^{\Phi_1} \preceq \omega_\pm^{\Phi_2}$ with $\Phi_1:=\Phi(t=1;s)$ to $\Phi_2:=\Phi(t=2;s)$, takes place if there exist at least one function $\psi(t; s)\in C^\infty_c(\mathbb{R})$ which is solution of the following differential equation
\begin{align}
\begin{aligned}
   \frac{d}{dt} \big( \overline{\Phi(t;s)} \Phi(t;u) \big)
   = &\ (i(1 - e^{s-u})+ {\alpha} e^s) \overline{\psi(t;s)} \psi(t;u)
   + {\alpha} e^{s} \big( \overline{\psi^\prime(t;s)} \psi(t;u) + \overline{\psi(t;s)}\psi^\prime(t;u)\big) .
\end{aligned}
    \label{condsuff}
\end{align}
where $\alpha = \pm 1$.

Let us discuss eqn. \eqref{condsuff}. \\
It appears that this equation is complicated to deal with. However, some properties can be obtained. First, it is easy to realize that constant $\Phi$ in the variable $t$, $\frac{d\Phi(t;s)}{dt}=0$, with $\psi(t;s)=0$ solve eqn. \eqref{condsuff}, which corresponds to stationary states. \\
Next, it is tempting to set $\Phi=\psi$ in \eqref{condsuff} and try to solve the resulting differential equation, which by setting formally $\rho(t;s,u)=\overline{\Phi(t;s)}\Phi(t;u)$ would take the form of a 2-dimensional transport equation. But this feature is only formal and does not seem unfortunately to be exploitable to solve the differential equation.\\

It is instructive to compare \eqref{condsuff} with its counterpart obtained in \cite{francwal22} from a slightly different Lorentzian Triple. This latter is characterized by a Dirac operator having a non-standard commutative ($\kappa\to\infty$) limit while the one considered in the present analysis satisfies $\lim_{\kappa\to\infty}\D\sim\slashed{\partial}$. Recall however that both Lorentzian Triples do not have a standard commutative limit so that the $\kappa\to\infty$ limit must be considered here as a pure mathematical manipulation.\\ Taking the large $\kappa$ limit of the differential equation \eqref{condsuff} is easily obtained by using \eqref{2d-derivation} and noticing that $s$ has mass dimension 1. By further setting $\psi=\Phi$, \eqref{condsuff} gives to the zero order in the $\frac{1}{\kappa}$ expansion
\begin{equation}
    \frac{d}{dt} \big( \overline{\Phi(t;s)} \Phi(t;u) \big)
   =  \big(i(u-s)+\alpha\big)\overline{\Phi(t;s)} \Phi(t;u)+\alpha\big( \overline{\Phi^\prime(t;s)} \Phi(t;u) + \overline{\Phi(t;s)}\Phi^\prime(t;u)\big).\label{Zecomparaison}
\end{equation}
which is identical to its counterpart found in \cite{francwal22} up to the multiplicative factors in the first term of the RHS which reduces to $i(u-s)$ in ref. \cite{francwal22}. This simply reflects the difference between the derivations used in both works. Indeed, the temporal derivation used in \cite{francwal22} is $P_0$ which is the large $\kappa$ limit of $X_0$ \eqref{2d-derivation}, i.e. $\lim_{\kappa\to\infty}X_0=P_0$ while the spatial derivation used in \cite{francwal22} simply vanishes at this limit.\\
Note that eqn. \eqref{Zecomparaison} is solved by any $\Phi$ solution of
\begin{equation}
   \frac{d}{dt} \Phi(t;s)=(is+\frac{\alpha}{2})\Phi(t;s)+\alpha \partial_s\Phi(t;s)\label{Zetransport},
\end{equation}
which is nothing but a transport equation. The solution is easily seen to have the form $\Phi(t;s)=\Phi_0(s+\alpha t)e^{t(is+\frac{\alpha}{2})}e^{\frac{i}{2}\alpha t^2}$ where $\Phi_0 \in C^\infty_c(\mathbb{R})$. Observe that the last factor $e^{\frac{i}{2}\alpha t^2} $ is unessential and can be simply omitted since it disappears from the defining relation of the pure states \eqref{etatpur} so that a continuous causal evolution between two pure states $\omega_\pm^{\Phi}$'s can occur whenever ${\Phi}=\Phi_0(s+\alpha t)e^{t(is+\frac{\alpha}{2})}$. Hence, at the zero order in $\frac{1}{\kappa}$, each evolution is characterized by an $\alpha t$ translation acting on $\Phi_0(s)$ combined with a temporal transformation represented by the rightmost exponential factor. As expected from the above discussion, this solution bears some similarity with the one found in \cite{francwal22}.

\section{Discussion.}\label{decadix}

This paper is the second of a stepwise exploration of the notion of causality in quantum (noncommutative) space-time $\kappa$-Minkowski described by Lorentzian Spectral Triples. In this analysis, we used a Lorentzian Triple differing from the one used in \cite{francwal22} by the Dirac operator having now the usual commutative limit. This choice singles out a particular abelian algebra of twisted derivations of the $\kappa$-Poincar\'e algebra.\\

As far as formal aspects are concerned, we have shown that the standard notion of Lorentzian Spectral Triples \cite{franc-eps2012}, \cite{franc2014} must be replaced by a twisted version in order to accommodate the twisted character of the derivations.
We have shown in particular that the twist appearing in (most of) the defining relations of the Lorentzian Triple is directly related to the twist of the twisted derivations. Note that the occurrence of a twist in the present Lorentzian Triple could have been {\it{a priori}} suspected as involving a Dirac operator similar to the one of a twisted Riemanian Spectral Triple, as discussed at the end of Subsection \ref{section23}.\\

As far as physical aspects are concerned, a central object in the description of causal structures modeled by a Lorentzian Triple is the causal cone. In this paper, we have exhibited relevant classes of functions belonging to the causal cone. In particular, analogs of the light-cone coordinates belong to these classes, see in particular \eqref{light-cone-coor}. These functions can be viewed as defining a quantum analog of the standard light-cone of the usual Minkowski space-time which (partly) encodes its causal structure. \\
Moreover, we have shown that these functions generate quantum constraints on the momentum and space coordinates as necessary conditions for a causal propagation between pure states to occur. Recall that pure states are the quantum analogs of points. In particular, causal functions of light-cone-type coordinates generate a quantum analog of the speed of light limit, see \eqref{limitvitesse}. \\

Let us discuss in physical words this latter constraint. First, use \eqref{chatdeschro1} and set $x_0=Ct$ where $C$ is a (dimensionfull) positive constant. Eqn. \eqref{limitvitesse} can then be recast into the form
\begin{equation}
     \frac{\vert\delta\langle X\rangle\vert}{\delta\langle t\rangle}\le C,\label{ehrenfest1}
\end{equation}
which is the Ehrenfest-type constraint to be fulfilled for two pure states, the analogs of points, to be causally related. In \eqref{ehrenfest1}, $\langle X \rangle$
denotes the mean value of the operator/observable $X$ in a given pure state, says $\phi_1$, which in a quantum mechanical framework may be interpreted as an average position for a system in the state $\phi_1$, while 
$\vert\delta\langle X\rangle\vert=\vert\langle X\rangle_{\phi_1}-\langle X\rangle_{\phi_2}\vert$ defines the distance between the two average positions for the states $\phi_1$ and $\phi_2$, which is some sense replaces the usual variation along a trajectory between two points in the commutative world. Recall that the notion of trajectory is meaningless in a noncommutative framework. A similar interpretation holds for $t$. \\
Thus, putting all together, it appears that the LHS of \eqref{ehrenfest1} may be interpreted as a velocity (holding for average quantities). Hence, \eqref{ehrenfest1} looks formally like the usual relation of special relativity $\frac{\delta x}{\delta t}\le c$ ($c$ is the speed of light) characterising the light-cone of the classical Minkowski space-time.\\
At this stage, some comments are in order.\\
First, note that a formally similar constraint has been found in \cite{francwal22} using a different Dirac operator. This may suggest that such a quantum analog of the speed of light limit is a feature of the quantum causality for $\kappa$-Minkowski space-time described by Lorentzian triples. Note that, leaving aside the technical (albeit important) questions of convergence and existence of causal functions, such a constraint stems formally from \eqref{evol-macro}, \eqref{evol-macro-bis}, independently of the Dirac operator.\\
It would be tempting to identify the constant $C$ in \eqref{ehrenfest1} with the speed of light. In this respect, the low energy (commutative) imprint of \eqref{ehrenfest1} would simply correspond to the usual relation $v\le c$ of special relativity. To verify carefully this conjecture needs however to properly approach the commutative limit starting from a Lorentzian triple and to control the deviations $\sim\mathcal{O}(\frac{1}{\kappa})$ from the classical situation. This will presumably give access to an estimate of the order of magnitude of effects related to the quantum light-cone linked to the Lorentzian triple and to confront them to those arising in the field theoretical approaches \cite{Neves_2010}, \cite{Mercati_2018a, Mercati_2018b}. This is not an easy task. It appears that the use of the GNS representation instead of the one used in the present work and in \cite{francwal22} seems to be a natural choice to reach this goal in a well controlled way and to extend the whole scheme to $3+1$ dimensions. This has been undertaken and will be presented elsewhere. Of course, twists will necessarily remain in the Lorentzian Triple. \\
Finally, some sufficient conditions for the existence of a continuous causal evolution between two pure states have been exhibited and examined and have been shown to be related to transport equations, see \eqref{Zetransport}, somewhat similar to the sufficient relations found in \cite{francwal22}.\\

The physical meaning (if any) of the constraints \eqref{mystery} remains obscure so far and we presently do no have a clear interpretation of these. Besides, note that we have only considered causal relations between pure states. The present study could be formally extended to non pure states but the precise physical interpretation of causal relations connecting non pure states (e.g. kind of wave packets) is not clear.\\

\section*{Acknowledgements}
\paragraph{}
K.H. and J.-C.W. thank the Action CA18108 QG-MM "{\it{Quantum gravity phenomenology in the multi-messenger approach }}" from the European Cooperation in Science and Technology (COST). J.-C. W. thanks P. Martinetti and T. Masson for fruitful discussions.

\appendix

\section{Basics on causality on a quantum space-time.}\label{ApendixA}
In the usual commutative framework, the notion of causality can be defined geometrically from points, curves and vectors.
Points are events in space-time. These are causaly related if they can be linked together by a causal curve. Recall that for a locally compact manifold, the
type of a curve is fixed by the corresponding nature of its tangent vectors which may be time-like, space-like, ... For instance, a causal
curve is a curve for which the tangent vectors at any point of the curve is time-like (or {null}). Whenever the manifold is globally hyperbolic, which is the case in the present paper,
this purely geometric description can be equivalently turned into an algebraic one. Indeed, it is known that points can be equivalently described as pure states{\footnote{Recall that a state of
an algebra is a positive linear functional with norm 1, $\varphi:\mathbb{A}\to\mathbb{C}$. For a manifold $M$, $\mathbb{A}=C^\infty_0(M)$, the algebra of smooth functions vanishing at infinity.}} while causal curves become causal functions, characterized as real-valued functions on $M$ which are non-decreasing along every future-directed causal curve. The set of causal functions forms, up to technical subtleties which are not essential here, a convex cone $\mathcal{C}$, a subset of the unitalization of $\mathbb{A}$, denoted by  $\widetilde{\mathbb{A}}$, which is called the {\it{causal cone}}.\\

One major result of \cite{cestbesnard} is that the causal cone determines entirely the usual causal structure of a globally hyperbolic manifold $M$ given by
\begin{equation}
p\preceq q\ \ \iff\ \ f(p)\le f(q),\ \ \forall f\in\mathcal{C}
\end{equation}
for any points/events $p,q\in M$. The next task is to characterize the causal cone in terms of algebraic objects. This has been carried out in \cite{franc-eps2012} using the data of the Lorentzian Spectral Triple modeling which is synthesized by \cite[Theorem 7]{franc-eps2012}. Indeed, the causal cone $\mathcal{C}$ can be equivalently defined as the convex cone of all real-valued functions $f\in\widetilde{\mathbb{A}}$ verifying
\begin{equation}
\forall \phi\in\mathcal{H},\\ \langle\phi,\mathcal{J}[D,\pi(f)]\phi\rangle\le 0,\label{cestzecone}
\end{equation}
where $\mathcal{H}$ is the Hilbert space entering the Lorentzian Spectral Triple, $D$ is the usual Dirac operator and $\mathcal{J}$ the fundamental symmetry. Then, \cite[Theorem 7]{franc-eps2012} guaranties that, for globally hyperbolic manifolds, the causal structure defined by
\begin{equation}
\forall \omega,\eta\in\mathfrak{S}(\widetilde{\mathbb{A}}),\ \omega\preceq \eta\iff \forall f\in\mathcal{C},\ \omega(f)\le\eta(f),\label{cestzecausalite}
\end{equation}
when restricted to the pure states in $\mathfrak{S}(\mathbb{A})$, the space of states, is {\it{exactly}} the usual causal structure of $M$.\\

The extension to a noncommutative framework of the above result is obvious and gives rise to \eqref{condcone1} and \eqref{cestcausal2bis}, which are linked to the data defining a Lorentzian Spectral Triple. This thus provides a description of a notion of causality on noncommutative space, which can be called quantum causality owing to the quantum nature of a noncommutative space.

\section{Relating derivations and representations.}
\label{apendix2}
\paragraph{}
For convenience, the computations will be performed with the mixed variables \eqref{reprerpmixed}. We also set $\beta:=\nu e^{-s}$, $\nu=-1,0,1$.\\

To prove \eqref{commut1}, we proceed as follows. The combination of \eqref{reprerpmixed} together with the relation
\begin{equation}
(\partial_0\phi)(s) = s \phi(s), \forall\phi\in\mathcal{H}_\nu\label{cestbe1}
\end{equation}
which transports the action of the initial self-adjoint operator $P_0$ to an action of a self-adjoint operator on any of the Hilbert spaces $\mathcal{H}_\nu$  \cite{CPT}, gives rise to the following equations $(\partial_0\pi_\nu(a)\phi)(s)=\int du\ s\widetilde{a}(u-s,\beta)\phi(u)$ and
$( \pi_\nu(a)\partial_0\phi)(s)=\int du\ u\widetilde{a}(u-s,\beta)\phi(u)$ which upon using $\partial_0\widetilde{a}(u) = iu \widetilde{a}(u)$ implies
\begin{equation}
    ([\partial_0,\pi_\nu(a)]\phi)(s)=(i\pi_\nu(\partial_0a)\phi)(s)\label{commut11}
\end{equation}
which holds for any $a\in\mathbb{A}_{x}$ and any $ \phi\in C^\infty_c(\mathbb{R})\subset\mathcal{H}_\nu$. This proves \eqref{commut1}. Note by the way that the self-adjointness of \eqref{cestbe1} w.r.t. the Hilbert product \eqref{hilbertproduct} is obvious.\\

Next, we prove \eqref{commut2}. The use of $X_0 = 1 - \mathcal{E}$, \eqref{cestbe1} and the first relation in \eqref{cestlaction} yields
\begin{equation}
    (X_0\phi)(s)=(1-e^{s})\phi(s),
    \label{act22}
\end{equation}
Note, {\it{en passant}}, the obvious self-adjointness w.r.t. the Hilbert product \eqref{hilbertproduct}. Eqn. \eqref{act22} is the counterpart of \eqref{cestbe1} for the operator $X_0$. Now, one computes
\begin{align}
    \big(\pi_\nu(\mathcal{E}(a))\phi\big)(s)
    = \int du\ e^{-(u-s)} \widetilde{a}(u-s, \beta) \phi(u),
\end{align}
so that the action of $\mathcal{E}$ amounts to a multiplication by $e^{- (u-s)}$. Then, one obtains
\begin{align}
    (X_0\pi_\nu(a)\phi)(s)
    &= \int du\ (1-e^{s})\widetilde{a}(u-s,\beta)\phi(u),\\
    \big( \pi_\nu(\mathcal{E}(a))X_0 \phi \big)(s)
    &= \int du\ \widetilde{a}(u-s,\beta)(e^{-(u-s)} - e^{s})\phi(u),
\end{align}
and again $\nu=-1,0,1$. Now, the twisted commutator \eqref{deftwist-commut} is defined as the difference of the two previous quantities so that
\begin{equation}
\begin{aligned}
    ([X_0,\pi_\nu(a)]_\mathcal{E}\phi)(s)
    &= \int du\ \big(1 - e^{s} - (e^{-(u-s)} - e^{s}) \big) \widetilde{a}(u-s,\beta)\phi(u) \\
    &= \int du\ (1 - e^{-(u-s)}) \widetilde{a}(u-s,\beta)\phi(u)
    = (\pi_\nu(X_0 a) \phi)(s),
\end{aligned}
\end{equation}
whihc leads to
\begin{equation}
   ([X_0,\pi_\nu(a)]_\mathcal{E}\phi)(s)=(\pi_\nu(X_0a)\phi)(s) ,\ \ \nu=-1,0,1\label{commut2bis}
\end{equation}
for any $a\in\mathbb{A}_{x}$ and any $ \phi\in C^\infty_c(\mathbb{R})\subset\mathcal{H}_\nu$.

\paragraph{}
Now, we turn to \eqref{commut3}. In the same way, one starts from
\begin{equation}
    (X_1\phi)(s) = i\nu e^s\partial_s\phi(s),\ \ \forall\phi\in\mathcal{H}_\nu
    \label{act222},
\end{equation}
which is obtained by combining \eqref{chatdeschro} and the expression for $X_1$. Note that the defined operator is no longer self-adjoint w.r.t the Hilbert product of $\mathcal{H}_\nu$. However, the final Dirac operator will stay self-adjoint, up to a twist, w.r.t. the Krein product, as required in Lorentzian Spectral Triples. Then, one computes
\begin{align}
    (X_1 \pi_\nu(a) \phi)(s)
    &= - i \nu\int du\ e^s \partial_a \widetilde{a}(u-s, \beta) \phi(u)
    - i \int du\ \partial_\beta \widetilde{a}(u-s,\beta) \phi(u), \\
    \big(\pi_\nu(\mathcal{E}(a)) X_1 \phi \big)(s)
    &= - i \nu\int du\ e^s \partial_a \widetilde{a}(u-s,\beta) \phi(u),
\end{align}
where an integration by part was used in the last equation. From this, one obtains the twisted commutator \eqref{deftwist-commut}
\begin{equation}
    ([X_1,\pi_\nu(a)]_\mathcal{E}\phi)(s)
    = - i \int du\ \partial_\beta \widetilde{a}(u-s,\beta) \phi(u)
    =  (\pi_\nu(X_1a)\phi)(s),
\end{equation}
for any $a\in\mathbb{A}_{x}$ and any $ \phi\in C^\infty_c(\mathbb{R})$.

\paragraph{}

\end{document}